\documentclass[conference]{IEEEtran}
\IEEEoverridecommandlockouts
\usepackage{amsthm,amsmath,amsfonts}
\usepackage{algorithmic}
\usepackage{algorithm}
\usepackage{array}
\usepackage[caption=false,font=normalsize,labelfont=sf,textfont=sf]{subfig}
\usepackage{textcomp}
\usepackage{stfloats}
\usepackage{url}
\usepackage{verbatim}
\usepackage{graphicx}
\usepackage{cite}
\usepackage{color,soul}
\usepackage{graphicx}
\usepackage{lipsum}
\usepackage{tabularx}
\usepackage{multicol}
\usepackage{booktabs}
\usepackage{multirow}

\usepackage[acronym,toc]{glossaries}
\newacronym{AoI}{AoI}{Age of Information}
\newacronym{AAoI}{AAoI}{Average \gls{AoI}}
\newacronym{AIRA}{AIRA}{Age Independent Random Access}
\newacronym{ADRA}{ADRA}{Age Dependent Random Access}
\newacronym{ADRA-MR}{ADRA-MR}{Age Dependent Random Access in Multiple Relays}
\newacronym{AIRA-MR}{\textsc{AIRA-MR}}{Age Independent Random Access in Multiple Relays}
\newacronym{ADRF}{ADRF}{Age Dependent Relay Forwarding}
\newacronym{CAP}{CAP}{Channel Access Probability}
\newacronym{CSI}{CSI}{Channel State Information}
\newacronym{DTMC}{DTMC}{Discrete Time Markov Chain}
\newacronym{LEO}{LEO}{Low Earth Orbit}
\newacronym{IoT}{IoT}{Internet of Things}
\newacronym{MTC}{MTC}{Machine Type Communications}
\newacronym{RA}{RA}{Random Access}
\newacronym{SIC}{SIC}{Successive Interference Cancellation}
\newacronym{SA}{SA}{Slotted ALOHA}
\newacronym{CSMA}{CSMA}{Carrier Sensing Multiple Access}
\newacronym{MIMO}{MIMO}{Multiple Input Multiple Output}
\newacronym{NOMA}{NOMA}{Non-Orthogonal Multiple Access}
\newacronym{ARQ}{ARQ}{Automatic Repeat Request}
\newacronym{SINR}{SINR}{Signal to Interference plus Noise Ratio}
\newacronym{LPWAN}{LPWAN}{Low Power Wide Area Network}
\newacronym{MR}{MR}{Multiple-Relay}
\newacronym{AD}{AD}{Age Dependent}
\usepackage[dvipsnames]{xcolor}

\def\BibTeX{{\rm B\kern-.05em{\sc i\kern-.025em b}\kern-.08em
    T\kern-.1667em\lower.7ex\hbox{E}\kern-.125emX}}
\begin{document}

\title{Age-of-Information Dependent Random Access in NOMA-Aided Multiple-Relay Slotted ALOHA\\
\thanks{This work has been supported in Brazil by CNPq (402378/2021-0, 305021/2021-4, 304405/2020-5), and in Finland by the Academy of Finland (6G Flagship No. 346208), and the Finnish Foundation for Technology Promotion.}}

\author{\IEEEauthorblockN{1\textsuperscript{st} Gabriel Germino Martins de Jesus}
\IEEEauthorblockA{\textit{Centre for Wireless Communications} \\
\textit{University of Oulu}\\
Oulu, Finland \\
gabriel.martinsdejesus@oulu.fi}
\and
\IEEEauthorblockN{2\textsuperscript{nd} João Luiz Rebelatto}
\IEEEauthorblockA{\textit{Department of Electronics} \\
\textit{ Federal University of Technology-Parana}\\
Curitiba, Brazil \\
jlrebelatto@utfpr.edu.br}
\and
\IEEEauthorblockN{3\textsuperscript{rd} Richard Demo Souza}
\IEEEauthorblockA{\textit{Department of Electrical and Electronical Engineering} \\
\textit{Federal University of Santa Catarina}\\
Florianópolis, Brazil \\
richard.demo@ufsc.br}
\and
\IEEEauthorblockN{4\textsuperscript{th} Onel Luis Alcaraz López} 
\IEEEauthorblockA{\textit{Centre for Wireless Communications} \\
\textit{University of Oulu}\\
Oulu, Finland \\
onel.alcarazlopez@oulu.fi}
}

\maketitle

\begin{abstract}
We propose and evaluate the performance of a Non-Orthogonal Multiple Access (NOMA) dual-hop multiple relay (MR) network from an information freshness perspective using the Age of Information (AoI) metric. More specifically, we consider an age dependent (AD) policy, named as AD-NOMA-MR, in which users only transmit, with a given probability, after they reach a certain age threshold. The packets sent by the users are potentially received by the relays, and then forwarded to a common sink in a NOMA fashion by {randomly selecting one of the available power levels, and multiple packets are received if all selected levels are unique}. We derive analytical expressions for the average AoI of AD-NOMA-MR. Through numerical and simulation results, we show that the proposed policy can improve the average AoI up to $76.6\%$ when compared to a previously proposed AD Orthogonal Multiple Access MR policy.
\end{abstract}


\section{Introduction}
The \gls{IoT} gives rise to many applications with performance requirements that current  technologies cannot fully meet~\cite{Nguyen:JIOT:2022,Popovski:IEEE:2022}. Some modern use cases with critical time requirements  suffer drastically if packets consistently fail to be delivered due to the overuse of the shared wireless channel. In that case, the so-called \textit{freshness} of information would not be guaranteed. In this light, quantifying the freshness of information becomes relevant, and the \gls{AoI} metric \cite{Kaul:INFOCON:2012,Kaul:ISIT:2017} can be used to this end. Notice that the freshest information is that with the lowest \gls{AoI}.

In order to comply with these and other requirements imposed by \gls{IoT} applications, many modern \gls{RA} policies have been proposed in recent years \cite{Clazzer:6Gsummit:2019}. Inspired by slotted ALOHA, 
several recent works \cite{Grybosi:JIOT:2022,Ren:WCNC:2022,Wang:TWC:2022} evaluate modern \gls{RA} methods from an \gls{AoI} perspective, including those designed to minimize the \gls{AoI} considering \gls{AD} schemes in  single- \cite{Atabay:INFOCOM:2020,Chen:ICCC:2020,Yavascan:2021:IJSAC,Chen:ITIT:2022,Ahmetoglu:IOTJ:2022} and two-hop networks~\cite{deJesus:IEEEAccess:2022}. Such extensive research has shown that the \gls{AoI} can be significantly reduced when the {transmission and reception parameters} in these methods are properly selected, inspiring further research in different setups.

In~\cite{Munari:3BCCN:2019, Munari:ITC:2021}, the authors consider a two-hop \gls{MR} network without direct links between users and destination, such that the communication is established through a set of dedicated relays. In~\cite{Munari:3BCCN:2019}, both the links between users and relays (the first hop) and between relays and destination (second hop) use slotted ALOHA, and the authors evaluate the influence of the number of relays in terms of throughput. A key observation is that while increasing the number of relays improves the performance of the first hop, it damages the second hop due to the increased number of collisions at the destination node. From an \gls{AoI} perspective, it is shown in~\cite{deJesus:IEEEAccess:2022} that adopting an \gls{AD} policy in the \gls{MR} scenario can reduce the network \gls{AAoI}, mainly by reducing the collision probability at the relays. However, the system performance is still limited by collisions in the second hop.       

In this contribution, we consider a similar \gls{AD}-aided two-hop \gls{MR} setup as in \cite{deJesus:IEEEAccess:2022}, but including the \gls{NOMA} capability in the second hop. In the so-called AD-NOMA-MR proposed scheme, the users can only start transmitting when their \gls{AoI} reaches a certain age threshold. The transmitted packets can potentially be captured by any of the available relays, which then forward the successfully received packets to the sink. The sink, in turn, is capable of retrieving more than one packet in a single time slot, as long as all of them are received in different power levels, by applying \gls{SIC}. {We derive analytical expressions to calculate the \gls{AAoI} of networks following this scheme, also showing}, through numerical and simulation results, that AD-NOMA-MR can significantly reduce the network \gls{AAoI} when compared to~\cite{deJesus:IEEEAccess:2022}, and that it approaches an ideal second hop, without packet loss, as the number of receive power levels grows.

\section{System Model and Preliminaries}

\begin{figure}[!t]
    \centering
    \includegraphics[width=0.5\textwidth]{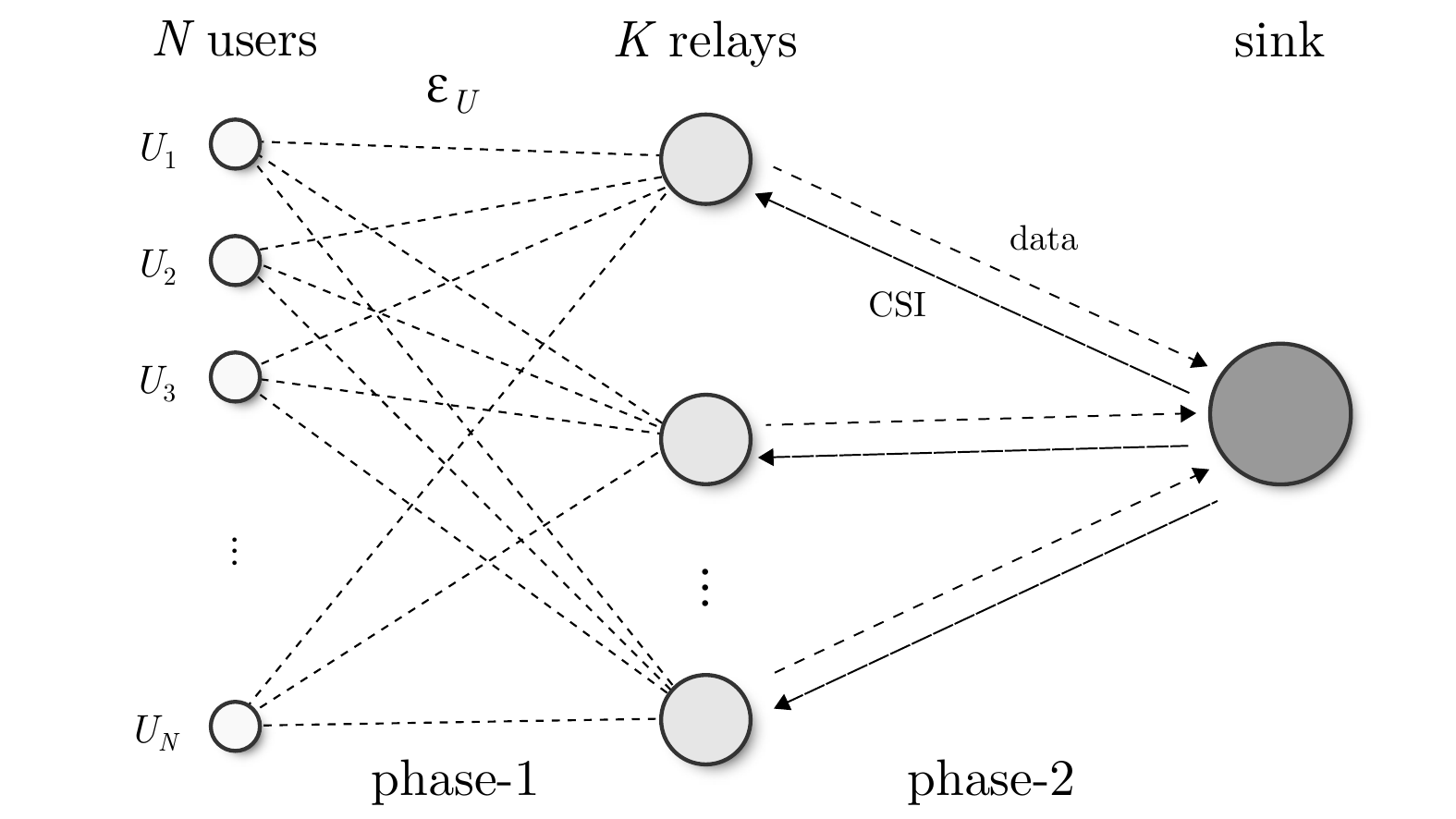}
    \caption{Illustration of the setup adopted in this work.}
    \label{fig:system_model}
\end{figure}

We consider an \gls{IoT} network composed of $N$ users denoted by $U_i, i=1,\dots, N$, which aim at transmitting their packets to a common destination (or sink). As in \cite{Munari:3BCCN:2019, Munari:ITC:2021, deJesus:IEEEAccess:2022}, it is assumed that there is no direct link between the users and the sink, such that the communication happens with the aid of $K$ relays in a two-hop fashion\footnote{Note that this setup is representative of some modern \gls{IoT} technologies, such as LoRaWAN \cite{LoRaWAN} and networks with \gls{LEO} satellites used as relays \cite{Qu:Access:2017}.}, as illustrated in Fig.~\ref{fig:system_model}.

Let us assume that time is divided in time slots of equal duration, and that a packet is generated, transmitted, received, and processed within a single time slot. In the first hop, which we refer to as {\it phase-1}, users generate data with probability $p$, a system design parameter subject to tuning, and transmit it as information packets through a shared wireless channel. We consider that the links between the users and relays are independent and identically distributed (i.i.d.), and packets transmitted in phase-1 are lost at the relays either due to erasure events (which happen with probability $\varepsilon_U$), or due to collisions with other packets transmitted in the same time slot and not erased at the considered relay. 

In the so-called {\it phase-2}, the second hop, relays immediately forward to the sink the packets they received from users. We assume that the relays have instantaneous \gls{CSI} of their channel to the sink. Note that this is a reasonable assumption since, in practice, $K$ will be typically much smaller than $N$, and the relays will likely be equipped with more powerful radios and processors compared to the users. Thus, a feedback link between these devices and the sink is feasible in practice, allowing for the phase-2 \gls{CSI} knowledge at the relays. In the case of channel reciprocity, this can be achieved by a pilot transmission from the sink to the relays at the start of each time slot. Upon the \gls{CSI} acquisition, the relays can adjust the transmit power to yield a given desirable power at the sink, operating under an erasure-free regime ({\it i.e.,} with erasure probability $\varepsilon_R=0$). However, despite the lack of erasures, collisions are still present in phase-2, as we do not assume packet scheduling among the relays.  Moreover, we assume that whenever a packet reaches the sink successfully, the user receives an acknowledgment before the start of the next time slot, so it can track its \gls{AoI}.

\subsection{Age-of-Information (AoI)} \label{ssec:aoi}
\begin{figure}[!t]
    \centering
    \includegraphics[width=0.4\textwidth]{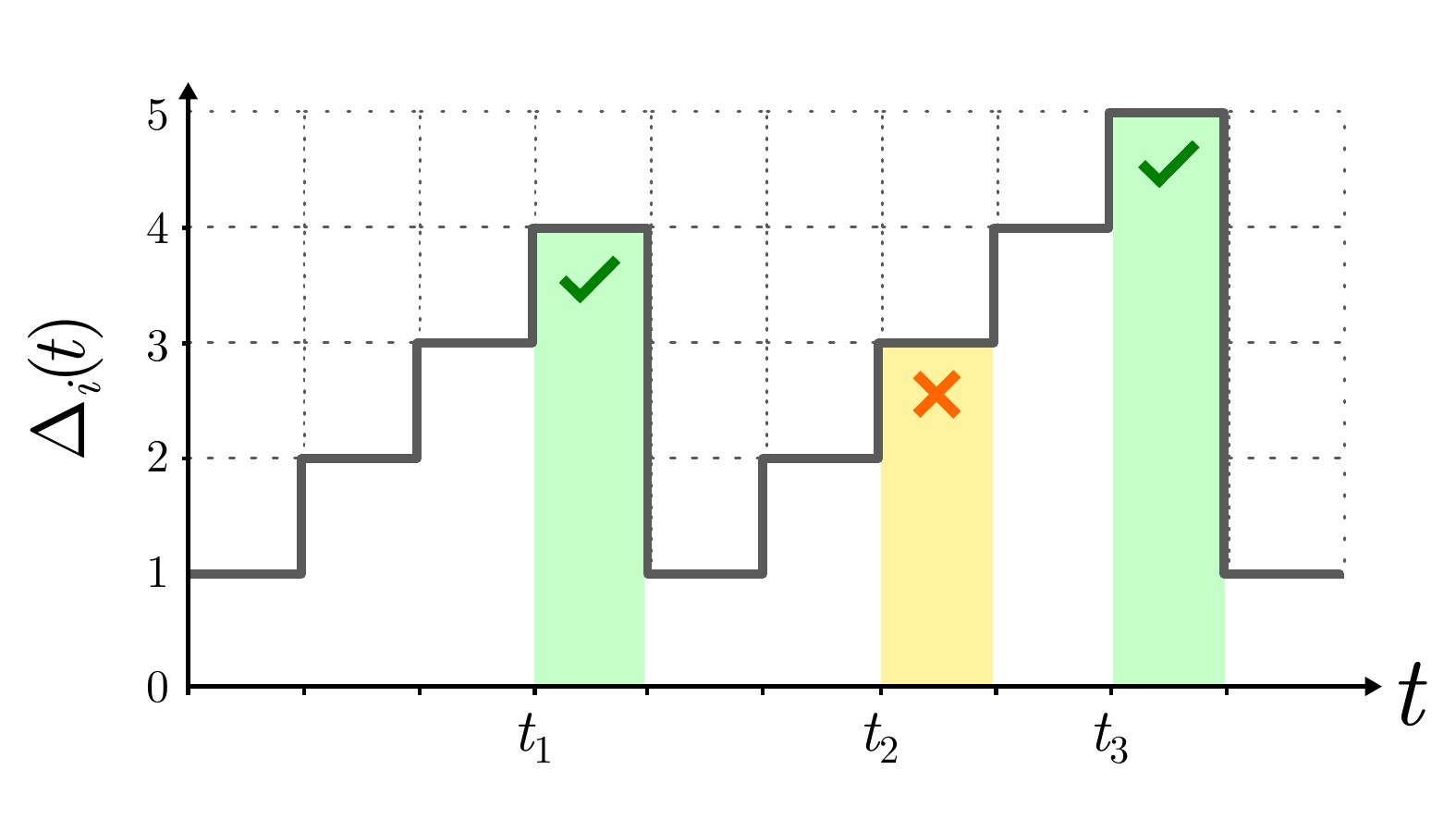}
    \caption{Example of the evolution of the \gls{AoI} of an user. This user transmits at $t_1$, $t_2$ and $t_3$. At $t_1$ and $t_3$, the packets are successfully received at the sink, resetting the \gls{AoI} to one. On the other hand, at $t_2$, the packet is lost, and its \gls{AoI} is increased by one in the subsequent time slot. }
    \label{fig:AoI_evo}
\end{figure}

The \gls{AoI} metric~\cite{Kaul:INFOCON:2012} quantifies the freshness of information by characterizing the time elapsed since the last packet received by the sink was generated. Let $r_i(t)$ be the generation time of the last packet from $U_i$ received by the sink at time slot $t$. The instantaneous AoI of the $i$-th user is a random process defined as $\Delta_i(t) = t - r_i(t)$~\cite{Kaul:ISIT:2017}. In a discrete-time system, at each time slot with no new information at the sink, a user's \gls{AoI} increases by one, and is reset to one otherwise, following a staircase shape, as illustrated in Fig.~\ref{fig:AoI_evo}. The time-averaged \gls{AoI}, or \gls{AAoI}, of $U_i$ is then obtained as~\cite{Chen:ICCC:2020}%
\begin{equation}\label{eq:avg_aoi_lim}
    \bar{\Delta}_i =  \lim_{T \rightarrow \infty} \frac{1}{T}\sum_{t=1}^{T}\Delta_i(t).
\end{equation}
The network \gls{AAoI} is determined by averaging the \gls{AAoI} of all devices as $
    \bar{\Delta} = \frac{1}{N}\sum_{i=1}^N \bar{\Delta}_i$.

\subsubsection{Age Dependent Random Access (ADRA)}
The authors from~\cite{Chen:ICCC:2020} propose a modern \gls{RA} method in which users comply with an \gls{ADRA} policy. This policy establishes that the users only attempt to transmit after reaching a certain age threshold $\delta$, and it is shown that, compared to regular \gls{SA}, the \gls{AAoI} decreases when $\delta$ and $p$ are properly selected, and there are optimal values for these parameters. Furthermore, the authors present a framework to calculate the network \gls{AAoI} by adopting an approximation to decouple the evolution of the \gls{AoI} of the users and model it as a discrete-time Markov chain. Let $q$ denote the probability of a packet from an user to be successfully delivered to the sink. Given that the probability of any user to have $\Delta_i(t)\geq\delta$ is $\theta= 1/(\delta pq + 1 - pq)$~\cite{Chen:ICCC:2020}, the network \gls{AAoI} becomes~\cite{Chen:ICCC:2020}
\begin{equation}\label{eq:average_delta}
    \bar{\Delta} = \frac{\delta}{2} + \frac{1}{pq} - \frac{\delta}{2(\delta pq + 1 - pq)}.
\end{equation}
Then, one can obtain from~\eqref{eq:average_delta} the \gls{AAoI} of a network in different setups by properly deriving an expression for $q$ in function of the system's parameters.

\subsubsection{ADRA in Multiple-Relays}

In~\cite{deJesus:IEEEAccess:2022}, the authors extend the \gls{ADRA} scheme from~\cite{Chen:ICCC:2020} to the \gls{MR} scenario from~\cite{Munari:3BCCN:2019} by proposing the so-called ADRA-MRU and ADRA-MRD schemes. The difference between such schemes lies in where the age-threshold principle from~\cite{Chen:ICCC:2020} is applied. Indeed, while in ADRA-MRU the users themselves only transmit when $\Delta_i(t)\geq\delta$ and the relays forward any packets they receive, in ADRA-MRD, the users transmit regardless of their instantaneous \gls{AoI} and the relays only forward the packets from the users with $\Delta_i(t)\geq\delta$. The results from~\cite{deJesus:IEEEAccess:2022} indicate that ADRA-MRU can often achieve lower \gls{AAoI} values compared to ADRA-MRD, due to its capability in reducing the number of collisions in phase-1, which is generally more congested. In order to evaluate the \gls{AAoI} of the networks in these schemes, the authors derive an expression for the probability of an user packet to be received successfully as given by  
\begin{equation}\label{eq:qADRAMR}
    q = \sum_{n=0}^{N-1}P_U(n)\left[K\hat{q}(1-(n+1)\hat{q})^{K-1}\right],
\end{equation}
where
\begin{equation}
    P_U(n) = \binom{N-1}{n}{(\theta p)}^n(1-\theta p)^{N-n-1},
\end{equation}
and $\hat{q} = (1-\varepsilon_U){\varepsilon_U}^n(1-\varepsilon_R)$ is the probability of a packet to be captured by a given relay when $n$ other packets transmit simultaneously. Notice that $\hat{q}$ is obtained as the product of {\it i)} the probability of the packet reaching a given relay without being erased while the remaining $n$ packets are erased, and {\it ii)} the probability of reaching the sink without being erased after being forwarded by the relay. If we consider an erasure-free phase-2 ($\varepsilon_R=0$), then \eqref{eq:qADRAMR} reduces to the success probability of ADRA-MRU in the system model adopted in this work.

Despite the erasure-free model in phase-2, the \gls{AAoI} obtained in~\eqref{eq:qADRAMR} still captures an interesting effect with respect to the number of relays: although increasing $K$ improves phase-1 performance (it becomes more likely that at least one relay will receive and forward a given packet), it also increases the congestion of phase-2, increasing collisions at the sink and rapidly compromising the overall system performance. Thus, there exists an optimal number of relays that minimizes the AAoI, which depends on the system parameters and is typically small for ADRA-MRU, {\it i.e.}, employing more relays than the optimal value does not improve (but damage, instead) the system performance~\cite{deJesus:IEEEAccess:2022}.  

Finally, it is worth mentioning that~\eqref{eq:qADRAMR} does not consider any sort of collision resolution at the sink, {\it i.e.}, no packet is recovered when multiple packets reach the sink simultaneously. In this paper, we build upon \eqref{eq:qADRAMR} to propose new random access policies in the \gls{MR} scenario. Specifically, we extend \cite{deJesus:IEEEAccess:2022} by introducing a \gls{NOMA}-aided policy in the next subsession.

\section{NOMA-Aided ADRA-MRU}

Herein, we propose an alternative to the ADRA-MRU scheme from~\cite{deJesus:IEEEAccess:2022} aiming at reducing the congestion in phase-2 and consequently boosting the spatial diversity benefits provided by the relays. In the so-called Age-Dependent \gls{NOMA} in Multiple-Relay (AD-NOMA-MR) proposed scheme, we consider that phase-2 employs power-domain NOMA~\cite{Yu:TWC:2021}, such that the sink possesses some collision resolution capability. More specifically, the proposed AD-NOMA-MR scheme works as follows:
\begin{itemize}
\item {\it Phase-1:} The CSI-unaware users {follow a generate-at-will policy, randomly sampling data and broadcasting their packets with probability $p$} only when $\Delta_i(t)\geq\delta$, as in ADRA-MRU~\cite{deJesus:IEEEAccess:2022}. The packets are not received by a relay in case of erasure (probability $\varepsilon_U$) or collision;
\item {\it Phase-2:} The relays that correctly received a packet in phase-1 forward it to the sink. To this end, they adjust their transmit power based on the available \gls{CSI} and randomly choose a power level among the set of $L$ available levels. Considering that the sink performs \gls{SIC}, the packets are not received in case of collision {\sl within the same power level}, {\it i.e.}, the sink is capable of resolving collisions as long as the collided packets employ different power levels~\cite{Yu:TWC:2021}.
 \end{itemize}
 
\begin{figure}[!t]
    \centering
    \includegraphics[width=0.5\textwidth]{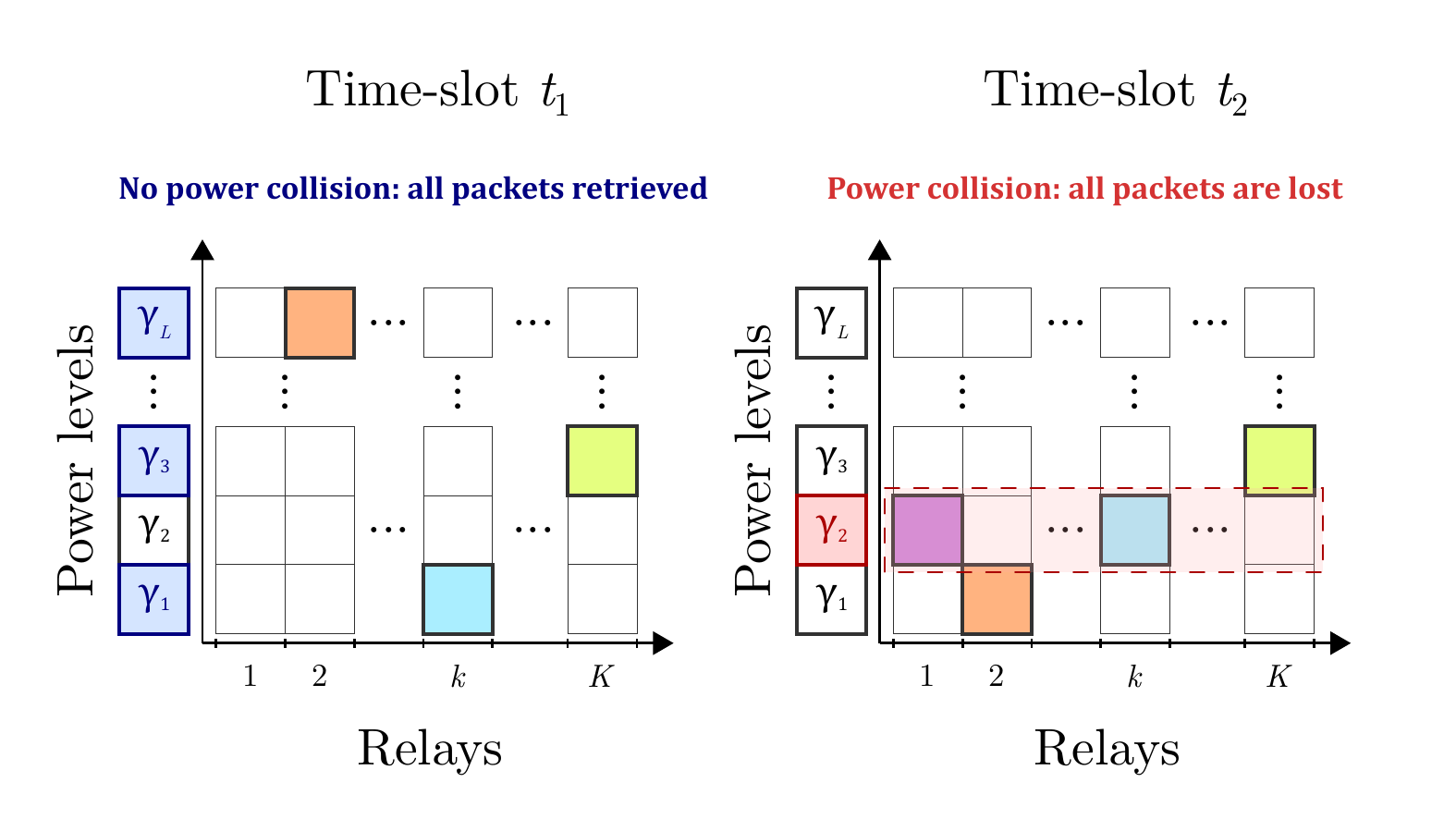}
    \caption{Illustration of collisions in the considered setup. At time slot $t_1$, all packets are successfully retrieved. At time slot $t_2$, there is a power collision which causes all packets to be lost.}
    \label{fig:ilustracao}
\end{figure}

Let $\gamma_l, l=1,2,\dots,L$ denote the $l$-th preset receive power level. Following~\cite{Yu:TWC:2021}, $\gamma_l=\Gamma\sigma^2(\Gamma+1)^{L-l}$, where $\Gamma$ is a target \gls{SINR} and $\sigma^2$ is the noise power, such that, after each iteration of the \gls{SIC} process, the resulting \gls{SINR} is at least equal to $\Gamma$. In this scheme, {when transmitting in power level $l$, the relays calculate the required transmit power such that the receive power at the sink is $\gamma_l$.} To this end, they rely on the available \gls{CSI}, which also guarantees that their packets will not be erased. At each time slot, the relays randomly select from any of these values, each value with probability $1/L$, and, if all transmitted packets arrive at the sink in different power levels, then all packets are successfully decoded. On the other hand, whenever more than one packet is received in the same power level, a power collision occurs and all packets transmitted in that time slot are lost, as now the target \gls{SINR} criteria  may not be met after each iteration. Fig.~\ref{fig:ilustracao} illustrates two possible outcomes of this approach: {\it i)} at time slot $t_1$ (figure on the left), all packets are correctly received, as there are no power collisions in any levels; while {\it ii)} at time slot $t_2$ (on the right), all packets are lost due to a power collision in power level $\gamma_2$. 

\subsection{Average AoI of AD-NOMA-MR} \label{ssec:aoi_AD-NOMA-MR}

We start our analysis by finding the probability of all packets transmitted by relays in a given time slot to be successfully decoded. In order for this to happen, all of the selected receive power levels must be different (and belong to the predefined set of $L$ power levels). Let $m$ relays transmit packets in a given time slot. There are $L!/(L-m)!$ permutations in which unique power levels can be distributed among $m$ relays, and the total number of permutations of power levels among the relays is $L^m$. Thus, the probability of all the selected levels to be unique is given by the ratio of these two values, as 
\begin{equation}\label{eq:PL}
    P_L(m) = \frac{L!}{L^m(L-m)!}.
\end{equation}

Next, we aim at finding the probability $q$ of a packet to be successfully delivered in the AD-NOMA-MR scheme. We recall that $\hat{q}$ is the probability that a packet from a user is captured by a relay and sent to the sink and that $n\leq N-1$ is the number of packets transmitted by other users in the same time slot. Then, the probability that $k+1\leq K$ relays forward the packets they have received to the sink, while the remaining $K-k-1$ relays do not, is given by
\begin{equation}
    \mathcal{P}_k(n) = \binom{K}{k+1} {\hat{q}}^{k+1}(1-(n+1)\hat{q})^{K-k-1},
\end{equation}
with the term $(1-(n+1)\hat{q})^{K-k-1}$ referring to the probability that all $n+1$ transmitted packets fail to be delivered through a subset with cardinality $K-k-1$ from all the $K$ relays available. Note that the $k+1$ relays can receive and forward any $(n+1)^{k+1}$ combinations of packets. However, we are interested in characterizing the success probability of the packet sent by a specific user. To this end, we define $\mathcal{N}_k(n)$ as the number of combinations in which the packet from this user is present. Then, this value is obtained by removing the number of combinations in which this packet is not received by any relays ($n^{k+1}$)  from the number of total combinations of packets received, as given by%
\begin{equation}
    \mathcal{N}_k(n) = (n+1)^{k+1}-n^{k+1}.
\end{equation}
From \eqref{eq:PL}, the probability of $k+1$ transmitters to select a distinct receive power level is given by $P_L(k+1)$. Furthermore, as one of the relays that captured the packet from the user will occupy one power level, for a successful reception without power  collisions, there can be at most $L-1$ other relays transmitting over phase-2. Thus, we can obtain the probability $Q$ of a packet to be successfully received at the sink when $n$ other packets are transmitted simultaneously by summing the product $\mathcal{N}_k(n) \mathcal{P}_k(n)$ over $k$ as
\begin{equation}\label{eq:P_sent}
    Q = \sum_{k=0}^{\min(L-1, K-1)}\mathcal{N}_k(n) \mathcal{P}_k(n) P_L(k+1).
\end{equation}
Finally, $q$ is obtained by summing $Q$ over $n$, multiplied by their probabilities $P_U(n)$, leading to
\begin{equation}\label{eq:qADNOMAMR}
    q = \sum_{n=0}^{N-1}P_U(n)\sum_{k=0}^{\min(L-1, K-1)}\mathcal{N}_k(n) \mathcal{P}_k(n) P_L(k+1).
\end{equation}

Note that \eqref{eq:qADNOMAMR} reduces to \eqref{eq:qADRAMR} when $L=1$, which we adopt as a baseline in the next section. Furthermore, one way to measure the full potential of this network topology is to consider phase-2 to be ideal, meaning that all packets retrieved by relays over phase-1 are successfully delivered to the sink. Although the practical feasibility of such a system is out of the scope of this work, the \gls{AAoI} performance under this hypothetical scenario serves as a lower bound to that of the proposed scheme. The probability $1-Q$ of a packet not reaching the sink for this case is readily given by the probability that all relays fail to capture and forward it, or $(1-\hat{q})^{K}$. Thus, $Q=1-(1-\hat{q})^{K}$, and 
\begin{equation}\label{eq:perfect_phase2}
    q = \sum_{n=0}^{N-1}P_U(n)(1-(1-\hat{q})^K).
\end{equation} 
Moreover, the same result can be obtained when $L\to\infty$ in \eqref{eq:P_sent}, as
\begin{equation}\label{eq:Q_perf}
\begin{split}
    Q & = \lim_{L\to\infty} \sum_{k=0}^{\min(L-1, K-1)}\mathcal{N}_k(n) \mathcal{P}_k(n) P_L(k+1)\\
      & =  \sum_{k=0}^{K-1}\mathcal{N}_k(n) \mathcal{P}_k(n)\lim_{L\to\infty}\frac{L!}{L^m(L-m)!}\\
        & =  \sum_{k=0}^{K-1}\mathcal{N}_k(n) \mathcal{P}_k(n)\lim_{L\to\infty}\frac{(L-1)(L-2)\dots(L-m+1)}{L^{m-1}}\\
        & =  \sum_{k=0}^{K-1}\mathcal{N}_k(n) \mathcal{P}_k(n)\lim_{L\to\infty} \sum_{i=1}^{m}\frac{s(m,i)L^{i-1}}{L^{m-1}}\\
        & =  \sum_{k=0}^{K-1}\mathcal{N}_k(n) \mathcal{P}_k(n)\left(1 + \lim_{L\to\infty} \sum_{i=1}^{m-1}\frac{s(m,i)}{L^{m-i}}\right)\\
      & = \sum_{k=0}^{K-1}\mathcal{N}_k(n) \mathcal{P}_k(n)\\
      & = \sum_{k=0}^{K-1}(n+1)^{k+1}\mathcal{P}_k(n) - \sum_{k=0}^{K-1}(n)^{k+1}\mathcal{P}_k(n),
\end{split}
\end{equation}
where $s(m,i)$ are the signed Stirling numbers of the first kind~\cite{Graham:book:1994}. Let us solve the two sum terms in \eqref{eq:Q_perf}, obtaining
\begin{subequations}
\begin{equation}\label{eq:sum1}
\begin{split}
    \sum_{k=0}^{K-1} (n+1)^{k+1} \mathcal{P}_k(n) = 1-\tilde{Q},
\end{split}
\end{equation}
\begin{equation}\label{eq:sum2}
    \begin{split}
    \sum_{k=0}^{K-1} n^{k+1} \mathcal{P}_k(n) =(1-\hat{q})^K-\tilde{Q},
\end{split}
\end{equation}
\end{subequations}
where $\tilde{Q}=(1-(n+1)\hat{q})^K$. Then, with \eqref{eq:sum1} and \eqref{eq:sum2},
\begin{equation}
\begin{split}
    \sum_{k=0}^{K-1} \mathcal{N}_k(n) \mathcal{P}_k(n) &=  1-\tilde{Q} - (1-\hat{q})^K + \tilde{Q} \\
    &=1-(1-\hat{q})^K,
\end{split}
\end{equation}which implies that the ideal phase-2 is a special case of AD-NOMA-MR when there is an infinitely large number of receive power levels available.

It is worth noting that achieving the required receive powers may not be possible for indiscriminately large numbers of power levels due to, for instance, limitations on transmit power and energy consumption. Reasonable values of power levels in practice should be relatively small.

\section{Numerical results}
\begin{figure}[!t]
    \centering
    \includegraphics[width=0.5\textwidth]{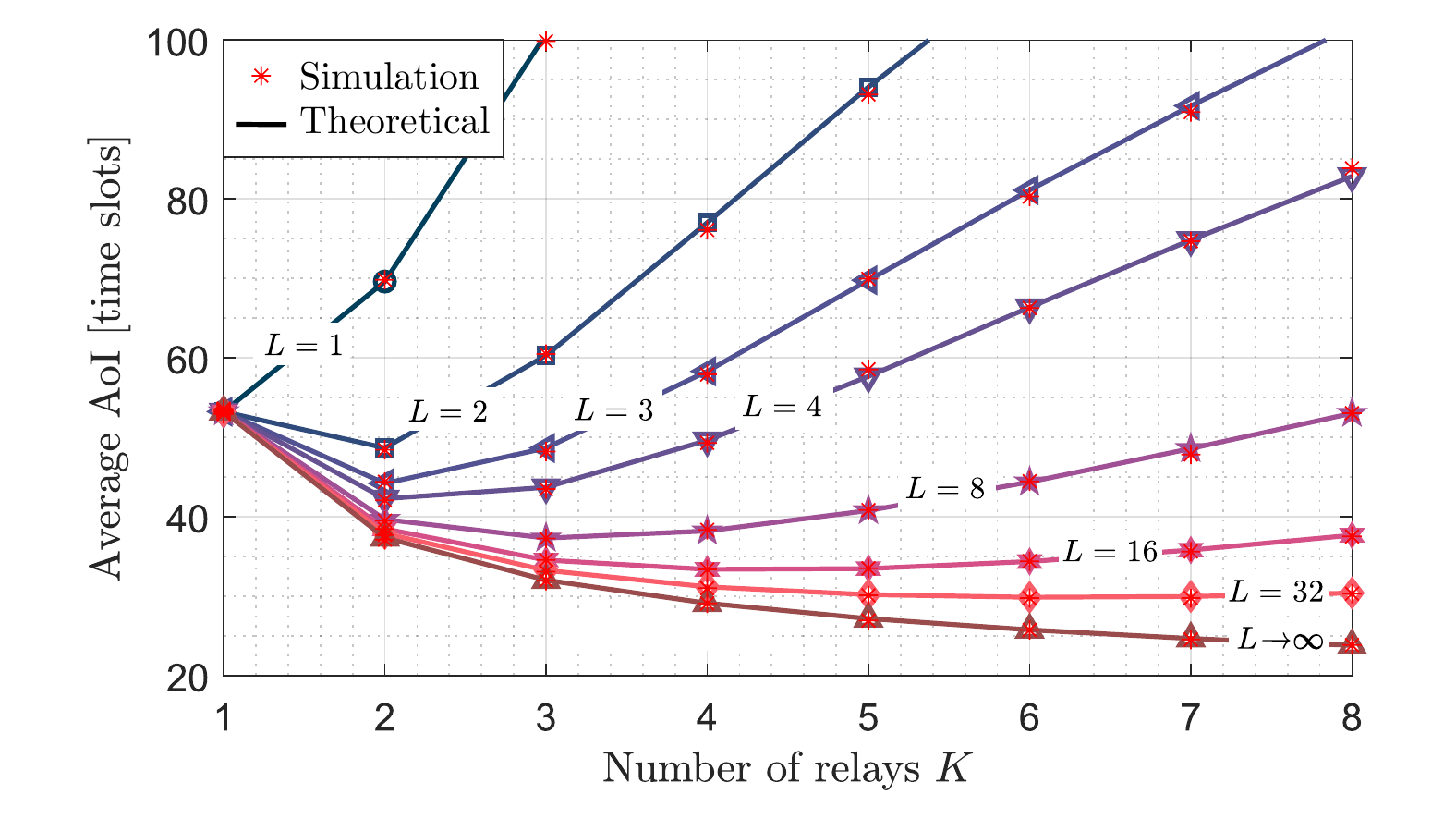}
    \caption{\gls{AAoI} versus $K$, for the AD-NOMA-MR scheme with different power levels $L$, $N=30$ users, phase-1 erasure rate $\varepsilon_U=0.3$ and numerically optimized values of $p$ and $\delta$ (see Table~\ref{tab:opt_delta}).}
    \label{fig:aoi_NOMA-MR}
\end{figure}

In Fig.~\ref{fig:aoi_NOMA-MR}, we present the network \gls{AAoI} under the AD-NOMA-MR policy for several values of $L$, as well as for the ideal phase-2 case. Moreover, Monte Carlo simulations are illustrated to validate the theoretical expressions derived in this work. We illustrate the performance of the proposed method by letting $N=30$ users transmit to $K\in[1,8]$ relays over a phase-1 characterized by $\varepsilon_U=0.3$. 
For each value of $L$ and $K$, the values of $p$ and $\delta$ are numerically optimized, such that the results presented are the lowest achievable \gls{AAoI} in each case. {We limit our search to $p\in[0,2/N]$~\cite{Chen:ICCC:2020} and $\delta\in[1,100]$. While in the numerical optimization the value of $p$ in each considered case converges to $p=0.067$}, the optimal values of $\delta$ vary and are listed in Table~\ref{tab:opt_delta}, with the overall optimal $(K,\delta)$ pair for each value of $L$ indicated in boldface. Notice that for $L=2$, 3, and 4, the optimal number of relays shifts from $K=1$ in the ADRA-MRU method to $K=2$, resulting in an \gls{AAoI} performance improvement with respect to ADRA-MRU of $9.04\%$, $20.4\%$, and $25.9\%$, respectively. For larger values of $L$, the optimal number of relays increases.  Specifically, the optimal values of $K$ for $L=8$, 16, and 32, are  3, 4, and 7, respectively. As expected, there is no optimal number of relays in the ideal phase-2 case, \emph{i.e.}, the AAoI decreases indefinitely as the number of relays increases.

\begin{table}[!t]
\caption{Optimized values of the age threshold $\delta$.}
\begin{center}
\begin{tabular}{ccccccccc} 
    \toprule
     \multirow{2}{*}{$L$} & \multicolumn{8}{c}{$K$} \\
        & $1$ & $2$ & $3$ & $4$ & $5$ & $6$ & $7$ & $8$ \\
        \midrule
        $1$                 & {\color{black}$\mathbf{47}$} &  $49$ & $46$ & $34$ & $18$ &   $1$ &   $1$ &  $1$ \\
        $2$                 & $47$ &  {\color{black}$\mathbf{38}$} & $37$ & $32$ & $24$ &  $15$ &   $7$ &  $1$ \\
        $3$                 & $47$ &  {\color{black}$\mathbf{35}$} & $32$ & $30$ & $24$ &  $18$ &  $12$ &  $7$ \\
        $4$                 & $47$ &  {\color{black}$\mathbf{34}$} & $30$ & $27$ & $24$ &  $20$ &  $14$ &  $9$ \\
        $8$                 & $47$ &  $32$ & {\color{black}$\mathbf{27}$} & $24$ & $22$ &  $20$ &  $17$ & $14$ \\
        $16$                & $47$ &  $31$ & $26$ & {\color{black}$\mathbf{23}$} & $20$ &  $19$ &  $17$ & $15$ \\
        $32$                & $47$ &  $31$ & $25$ & $22$ & $19$ &  $18$ &  {\color{black}$\mathbf{16}$} & $15$ \\
        $\to\infty$         & $47$ &  $31$ & $24$ & $21$ & $18$ &  $16$ &  $15$ & {\color{black}$\mathbf{14}$} \\
    \bottomrule
\end{tabular}
\end{center}
\label{tab:opt_delta}
\end{table}

\begin{figure}[!t]
    \centering
    \includegraphics[width=0.45\textwidth]{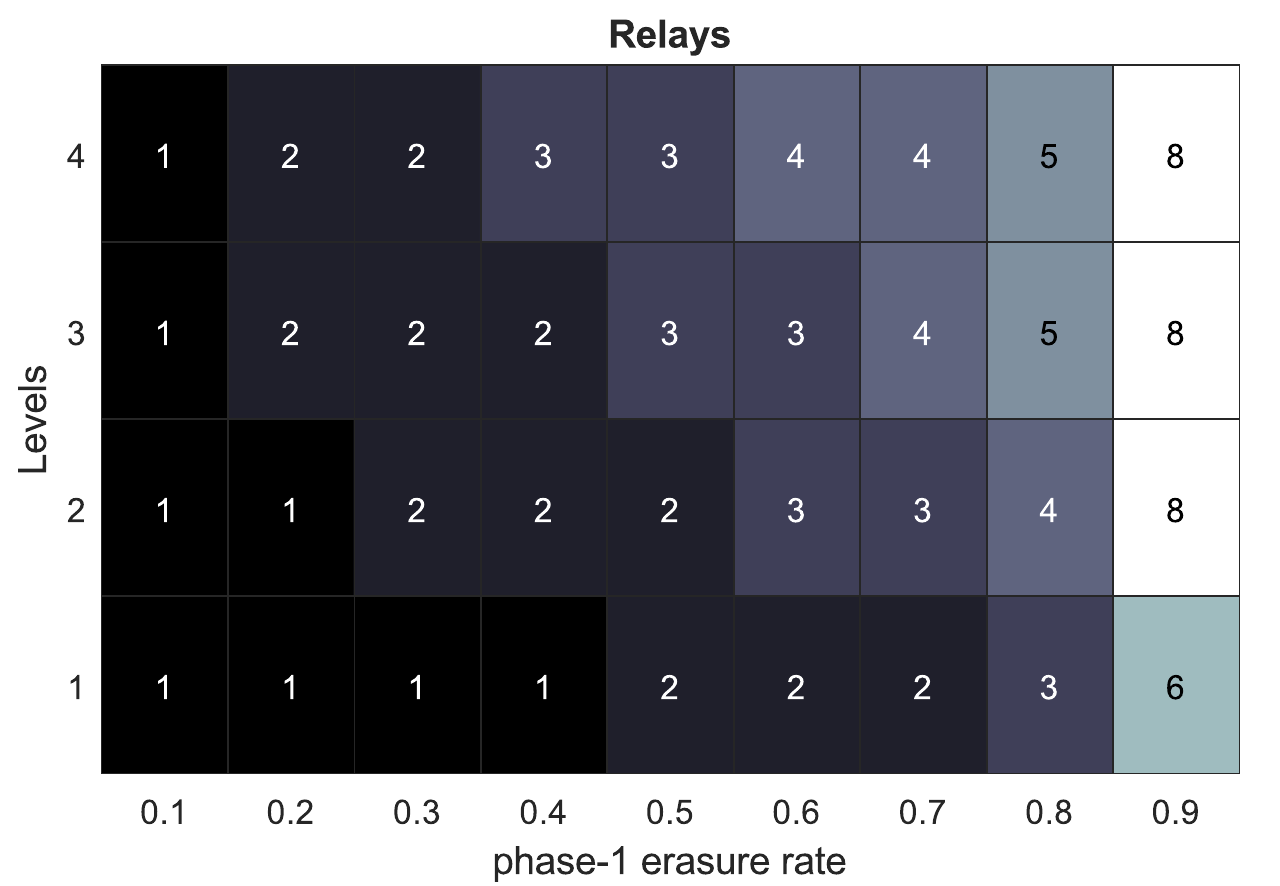}
    \includegraphics[width=0.45\textwidth]{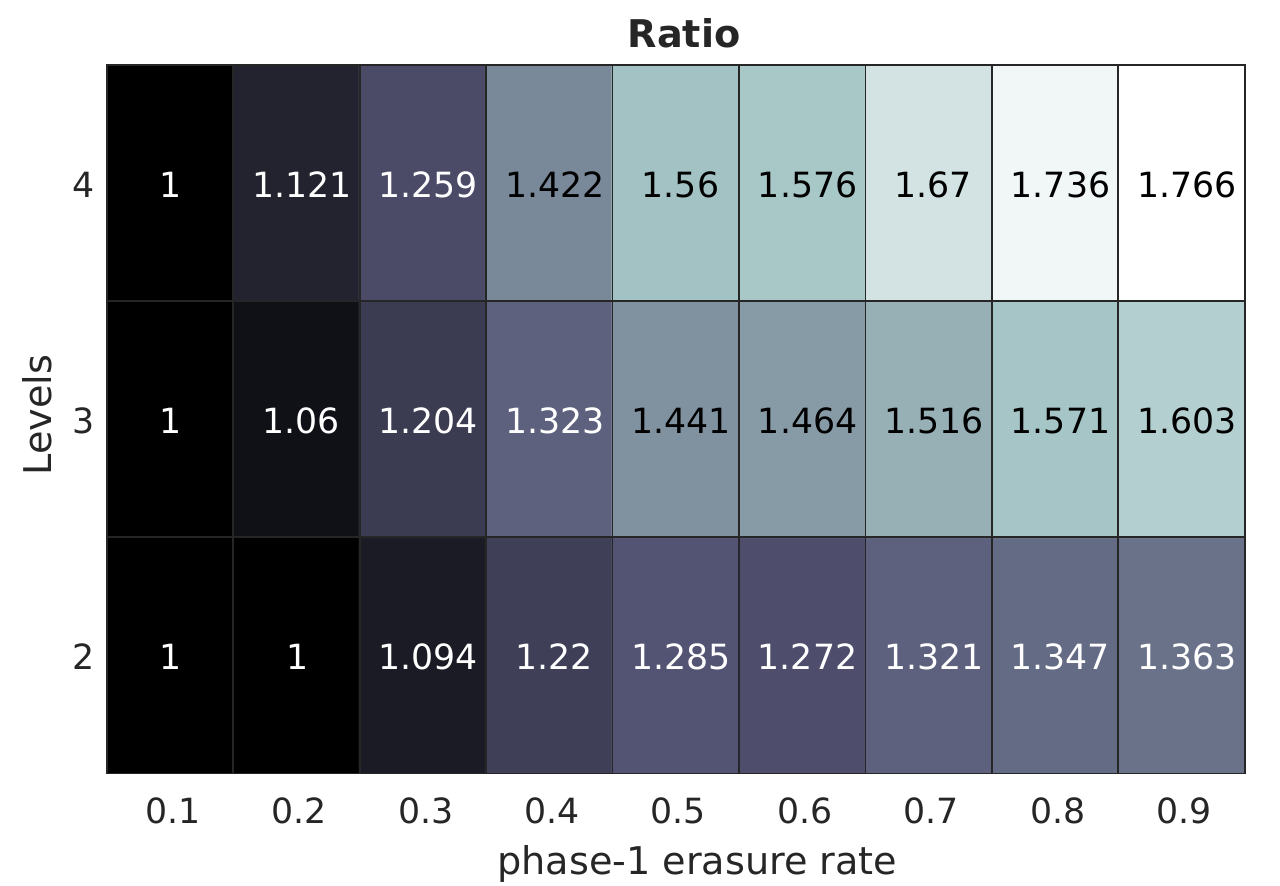}
    \caption{Optimal $K$ (top) and the ratio of the \gls{AoI} obtained in the ADRA-MRU (with no erasure on the phase-2) policy and AD-NOMA-MR policy (bottom) with $L\in[2,4]$, as the quality of the channels deteriorate. }
    \label{fig:aoi_NOMA-MR-ratio}
\end{figure}

Increasing the number of receive power levels $L$ increases the power consumption at the relays. Even if  the relays are powered by external sources other than batteries, these devices will still be subject to power limitations due to either local regulations or hardware constraints, which makes it of practical relevance to evaluate this scheme for small numbers of power levels. Thus, in Fig.~\ref{fig:aoi_NOMA-MR-ratio}, we present the optimal number of relays (top), and the performance gain (bottom) under AD-NOMA-MR with $L\in[2,4]$ compared to $L=1$, for several values of phase-1 erasure probability $\varepsilon_U$. Note the significant performance improvement as the number of power levels increases and the quality of phase-1 channel deteriorates. As the channel quality deteriorates, more relays are needed to successfully retrieve packets in phase-1. In turn, as more relays are available, the probability of more than one relay capturing a packet increases. When this happens, the availability of several power levels allows for the reception of these packets, resulting in this significant performance gain.

\section{Conclusion}
In this paper, we extended the ADRA-MRU policy by including the possibility of \gls{NOMA} in the second hop, which we called the AD-NOMA-MR policy. Specifically, the users transmit packets to a set of relays with a given access probability after reaching an \gls{AoI} threshold. The relays then forward the packets to a sink while targeting a specific receive power level. If all packets arrive at the sink with different power levels, they can be successfully retrieved, otherwise no packet is recovered. {In order to evaluate the performance of these networks, we derived expressions to calculate the \gls{AAoI}} and numerically optimized the access probability of users, age threshold, and number of available relays for several values of receive power levels. We showed that the proposed method is capable of reducing the network \gls{AAoI} in up to $76.6\%$ compared to ADRA-MRU.
\bibliographystyle{IEEEtran}
\bibliography{references}

\end{document}